\documentclass[UTF8]{article}
\usepackage{spconf,amsmath,graphicx}
\usepackage{hyperref}
\usepackage{makecell,booktabs}

\usepackage{enumitem}
\usepackage{tabularx}
\usepackage{url}
\usepackage{multirow}

\title{Expressive TTS Driven by Natural Language Prompts Using few Human Annotations}
\name{Hanglei Zhang, Yiwei Guo, Sen Liu, Xie Chen, Kai Yu\sthanks{Corresponding author.\\  © 2024 IEEE. Personal use of this material is permitted. Permission from IEEE must be obtained for all other uses, in any current or future media, including reprinting/republishing this material for advertising or promotional purposes, creating new collective works, for resale or redistribution to servers or lists, or reuse of any copyrighted component of this work in other works.}}
\address{MoE Key Lab of Artificial Intelligence, AI Institute\\X-LANCE Lab, Department of Computer Science and Engineering\\
Shanghai Jiao Tong University, Shanghai, China\\
\texttt{op.131@sjtu.edu.cn}}
\begin{document}
\ninept
\maketitle
\begin{abstract}

Expressive text-to-speech (TTS) aims to synthesize speeches with human-like tones, moods, or even artistic attributes. 
Recent advancements in expressive TTS empower users with the ability to directly control synthesis style through natural language prompts. 
However, these methods often require excessive training with a significant amount of style-annotated data, which can be challenging to acquire. 
Moreover, they may have limited adaptability due to fixed style annotations.
In this work, we present FreeStyleTTS (FS-TTS), a controllable expressive TTS model with minimal human annotations.
Our approach utilizes a large language model (LLM) to transform expressive TTS into a style retrieval task.
The LLM selects the best-matching style references from annotated utterances based on external style prompts, which can be raw input text or natural language style descriptions. 
The selected reference guides the TTS pipeline to synthesize speeches with the intended style.
This innovative approach provides flexible, versatile, and precise style control with minimal human workload. 
Experiments on a Mandarin storytelling corpus demonstrate FS-TTS's proficiency in leveraging LLM's semantic inference ability to retrieve desired styles from either input text or user-defined descriptions. This results in synthetic speeches that are closely aligned with the specified styles.
\end{abstract}


\section{INTRODUCTION}
\label{sec:intro}
Although neural text-to-speech (TTS) models like~\cite{FS2,VITS,VQTTS} can achieve substantial improvements in the quality and naturalness of synthesized speech, there remains a notable gap in the expressiveness between synthetic and human speech, which involves the complex manipulation of style-related features such as tones, sentiments, etc.
This limits TTS systems to cater to diverse scenarios and application needs, giving birth to the research of expressive TTS. 

An expressive TTS system is tasked with synthesizing speeches in accordance with specific styles of explicit style descriptions or implicit sentiments within the content.
Efforts have been made to enable the control of explicit style descriptions in expressive TTS. Prior methods relied on supervised approaches with predefined sets of style classes, such as emotions~\cite{classes1,um2020emotional,guo2023emodiff}, to guide expressive synthesis. Besides, continuous style spaces allowing for flexible control beyond predefined categories are more practical.
These methods typically employ a reference encoder~\cite{GST} to extract style embeddings unsupervisedly from reference audios serving as synthetic guidance. Subsequently, various techniques upon this approach have been developed ~\cite{GST,Global-VAE,latent,du2021phone,guo2022unsupervised} to obtain better control ability. 
However, these methods are still limited to cater to diverse style prompts in natural language format, as the unsupervised latent space is not intuitive and less interpretable. 

Addressing this is crucial because natural language is more user-preferred and precise in conveying human intentions. Thus, recent works~\cite{PromptTTS,ST-TTS,InstructTTS} try to develop expressive TTS directly driven by natural language style prompts. They explicitly align the style embeddings from annotated references with the semantic embeddings of the annotations extracted by BERT-based models~\cite{BERT,RoBERTa} in a cross-modal latent space. While these approaches are closer to natural-language-driven systems, they exhibit significant shortcomings, especially the considerable workload required for annotation and training, as well as a decrease in performance when dealing with user inputs beyond predefined annotation scenarios.

To address these issues, we introduce FreeStyleTTS (FS-TTS), a novel expressive TTS strategy driven by natural language prompts with minimal human annotations. Our approach distinguishes itself from the aforementioned methods by reframing the expressive TTS problem as a style retrieval task, rather than a cross-modal alignment task. This shift allows us to reduce the need for extensive annotations significantly.
Specifically, we build a TTS model with a reference encoder based on a variational autoencoder (VAE) for style modeling and manually annotate the natural language style description for fewer than 300 audio utterances. Subsequently, we employ a large language model (LLM) to retrieve the best-matched reference style from the annotated utterances based on the user's style instruction or the input text itself. This selected reference then guides the TTS model in synthesizing speech of the desired style. We conduct experiments on a highly expressive Mandarin storytelling dataset to validate the effectiveness and controllability of the proposed system.

In short, our work contributes to building an expressive TTS system with a powerful style control, as well as an easier annotation process and less annotation workload without loss of generation quality for diverse natural language prompts and texts. It is therefore a step closer to the ideal user-controllable expressive TTS system. Audio
samples are available online
\footnote{\url{http://hlz1012.github.io/FS-TTS}}

\section{FreestyleTTS with LLM style retrieval}
\label{sec:format}

\begin{figure}[htb]
\begin{minipage}[b]{1.0\linewidth}
  \centering
  \centerline{\includegraphics[width=7.8cm]{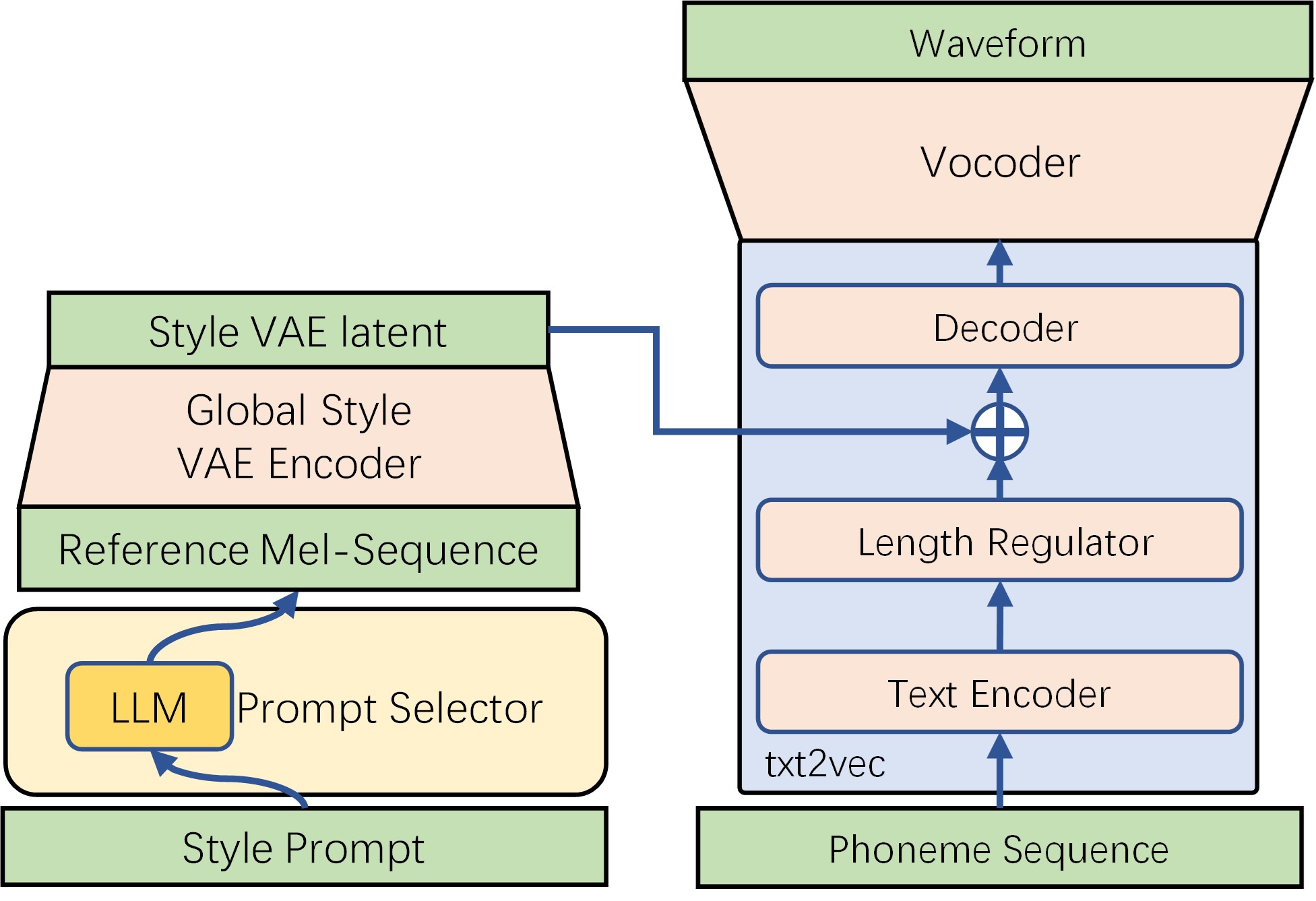}}
\end{minipage}
\caption{The architecture of FreeStyleTTS, with Global Style VAE Encoder, LLM Prompt Selector and TTS pipeline}
\label{Model}
  \vspace{-0.4cm}
\end{figure}

The overview of FreeStyleTTS is illustrated in Fig.\ref{Model}. The whole architecture consists of a Global Style VAE Encoder, an LLM Prompt Selector, and a TTS pipeline (txt2vec and vec2wav). Our entire system focuses on style control in the single-speaker speech synthesis scenario. It works by taking a given natural language style prompt, retrieving the corresponding reference from our annotations using the LLM Prompt Selector, and controls the TTS pipeline for speech synthesis by the reference's VAE latent variable. We'll discuss them and our strategy in detail.
\vspace{-0.2cm}
\subsection{Global Style VAE Encoder}
The style embeddings of the reference utterances are extracted by the Global Style VAE Encoder, following the structure proposed by Zhang et al. \cite{Global-VAE}. This approach introduced VAE~\cite{VAE} into the unsupervised speech style representation for the first time, enhancing the early Global Style Token~\cite{GST} in terms of style control and transfer capabilities. The input encoder network comprises six 2-D convolutional layers, a ReLU activation, and a final GRU layer, closely resembling the Reference Encoder introduced in the aforementioned works.
The only difference is that we remove the batch-normalization layers to make sure the outputs are distinguishable enough for expressiveness. As described in \cite{Global-VAE}, two
separate fully connected layers project the output embeddings into the means $\boldsymbol{\mu}$ and log-scale standard deviations  $\boldsymbol{\sigma}$ of the VAE latent variables. 
Employing reparameterization, we sample a $\mathbf{z}'$ from standard Gaussian distribution and calculate the VAE latent variable $\mathbf{z} =\boldsymbol{\mu}+\mathbf{z}'\odot\boldsymbol{\sigma}$.

To avoid the degeneration of controllability and synthesis diversity resulting from overfitting or KL-vanishing, we employ the closed-form KL-divergence loss for Gaussian distribution and utilize KL-annealing~\cite{KL-An} to ensure a relatively continuous and informative distribution of the latent variables.

\begin{figure}[h]
\begin{minipage}[b]{1.0\linewidth}
  \centering
  \centerline{\includegraphics[width=6.7cm]{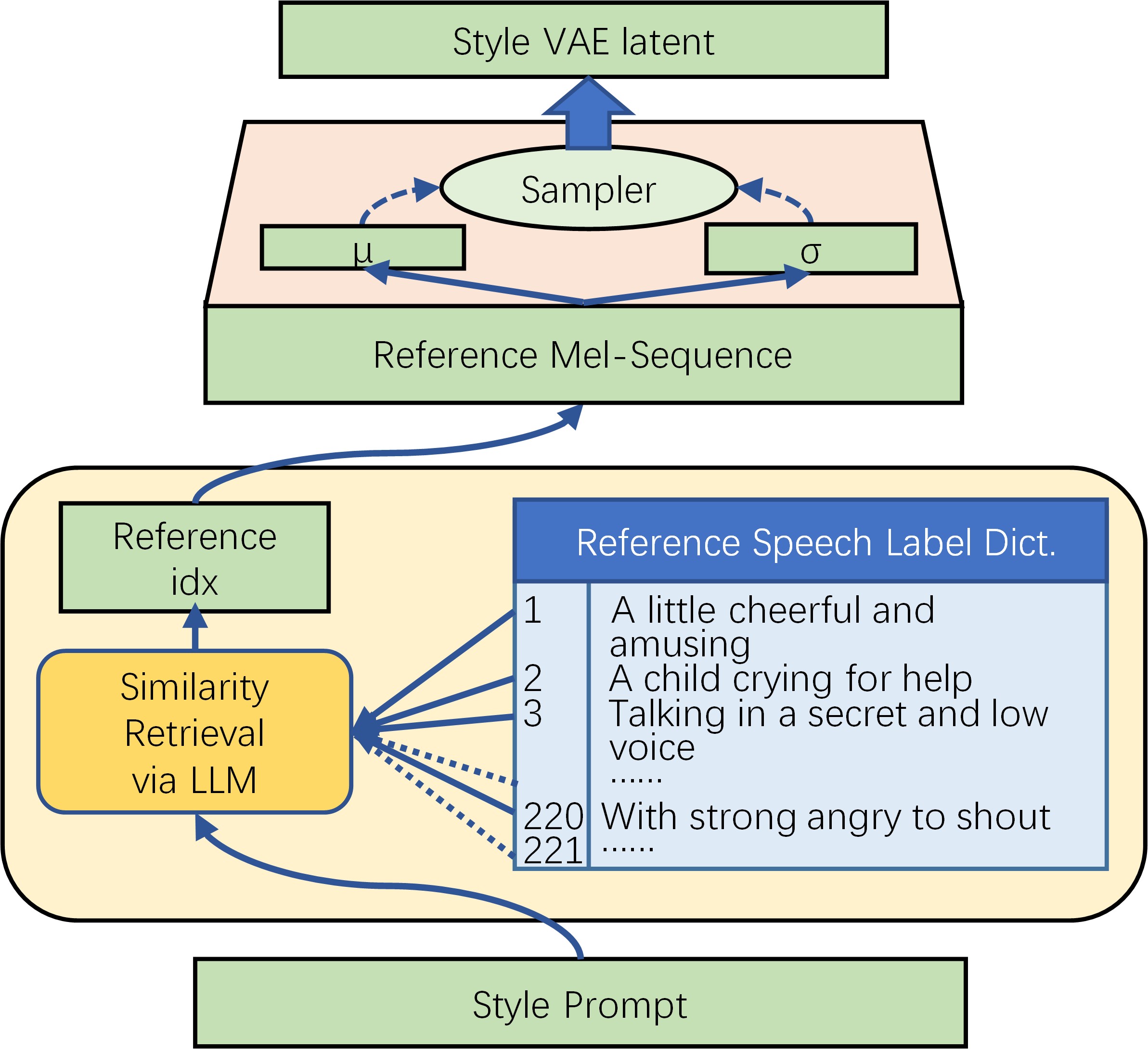}}
\end{minipage}
\caption{LLM Prompt Selector and Global Style VAE Encoder in detail}
\label{VAE}
  \vspace{-0.4cm}
\end{figure}
\vspace{-0.2cm}

\subsection{LLM Prompt Selector and Annotation Strategy}
\label{2.2}
One challenge in the earlier reference style control systems is the selection of the best-matching reference based on users' specified style description or the semantic content within the input. As mentioned before, instead of creating a prompt-speech style latent space, which would require a significant amount of annotating and training effort, we directly connect natural language prompts and references through LLM with few annotations. Our approach can be described by the following style retrieval stages:

\begin{enumerate}[leftmargin=*,noitemsep, topsep=5pt]
    \item Pick a moderate number of utterances randomly and label them with style descriptions in natural language without a fixed format. 
    Specifically, we describe each reference's style with 1-4 words or phrases, which can be any adjectives about the speaking scenario, emotion, loudness, pitch and speaking rate, etc.
    \item Store the labels in a dictionary as key-value pairs. eg. \{(1,"A serious and boastful tone"),(2, "A striking tone with highly dynamic in volume.")$\cdots$\}. The index keys and reference items are in one-to-one correspondence.
    \item Let LLM function as the selector. For style control scenarios, the LLM is prompted by "pick the label in the dictionary best matching with our given description of speech styles, our dictionary is \{$\cdots$\}", followed by our required styles or emotions in natural language.
    Alternatively, in style inference scenarios where the TTS system needs to synthesize with the style according to the input text, we also directly input the sentence into the LLM selector and ask it to return the best matched style reference. 
    \item In implementation, the index of that reference is returned by the LLM selector to locate the mel-sepctrogram of the best matched reference. Then this reference mel-spectrogram is fed into the style encoder for further control on the TTS acoustic model.
\end{enumerate}
The process is illustrated on the left part of Fig.\ref{Model}.
Also, Table \ref{Anno} shows some examples of the inputs and LLM selector's outputs. 


\begin{table}[ht]
   \centering
   \scriptsize
    \vspace{-0.4cm}
   \caption{Prompt Examples}
   \begin{tabularx}{0.49\textwidth}{XXX}
   \toprule 
  \multicolumn{1}{c}{\textbf{\vtop{\hbox{\strut  \textbf{\footnotesize Style description}}\hbox{\strut  \textbf{\footnotesize from user}}}}} & \multicolumn{1}{c}{\textbf{\vtop{\hbox{\strut  \textbf{\footnotesize Sentence}}\hbox{\strut  \textbf{\footnotesize for synthesis}}}}} & \multicolumn{1}{c}{\vtop{\hbox{\strut \textbf{\footnotesize Reference style}}\hbox{\strut  \textbf{\footnotesize retrieved by LLM}}}} \\ \midrule
   I fell into the water and shouted for help & ``Hurry up! Somebody call an ambulance!" & The tone of a shrill voice and an urgent cry for help \\ \midrule
   I whispered conspiracy. & ``Shh, we should sneak through the room." & Speaking privately with a speculative tone \\ \midrule
   Complaining sadly with a sense of frustration. & ``Too late, my days are numbered." & Somewhat weary and melancholic \\ \midrule
   Bragging proudly about himself. & ``Mom, I got A+ in the final test!" & In a triumphant, proud tone \\ \bottomrule
   \end{tabularx}
   \label{Anno}
   \vspace{-0.2cm}
\end{table}

In this way, our strategy handles both the explicit prompt of specified style description and the implicit prompt from the semantic content of the text to be synthesized. 
The former performs direct comparison and retrieval on semantic similarity, which is also called a style selection task; while the latter performs semantic sentimental analysis, i.e. the style inference task.

Our annotation method differs from the previous methods in three following perspectives:
\begin{enumerate}[leftmargin=*,noitemsep, topsep=5pt]
\item \textbf{Fewer restrictions and requirements}. Unlike fixed words in PromptTTS \cite{PromptTTS}, our labels are closer to the free natural language style description in the proprietary dataset in InstructTTS~\cite{InstructTTS}.
By the powerful semantic inference ability of LLMs, the style description labels can have any form in natural language.
\item \textbf{Easy extensibility}. The labels in this framework are not fixed within the dataset, so users can choose other unlabelled references or change the style descriptions according to their intention without training again and change the whole style latent space.
For other approaches with shared latent space for both prompts and audio, such flexibility is hard to obtain. 
\item \textbf{Very small in amount}. Compared with thousands of labels in the expressive datasets in previous works, the labeling procedure in this framework does not require much labor force. And
out-of-domain natural language prompts can still be treated as in-domain for LLMs, whose effectiveness has been proved~\cite{LLM1,LLM2}.
Together with the sample diversity brought by the VAE, a limited amount of labels are enough to provide rich expressiveness.
\end{enumerate}
In the following sections, detailed ablations on the number of annotations and LLM configurations will be presented.
 \vspace{-0.2cm}
\subsection{TTS Pipeline}
Our TTS pipeline is based on VQTTS~\cite{VQTTS}, a high-fidelity TTS system using self-supervised vector-quantized (VQ) acoustic features rather than mel-spectrograms. 
The acoustic model txt2vec first encodes the input phones into hidden states $\boldsymbol{h}$ with a text encoder, and then predicts duration per phone from $\boldsymbol{h}$. 
After that, $\boldsymbol{h}$ is repeated according to the durations, and the repeated style latent is projected and added onto it. 
A decoder finally predicts the VQ features from a pre-trained vq-wav2vec~\cite{vqw2v} together with prosody features including log pitch, energy, and probability of voice~\cite{Kaldi}. 
The vocoder vec2wav converts these features into waveforms. 
The training configuration of vec2wav is the same as VQTTS, while the criterion for text2vec is slightly different:
\begin{equation}\label{eq:loss}
    \mathcal{L}_{\texttt{txt2vec}} = \mathcal{L}_\texttt{dur} + \mathcal{L}_\texttt{VQ}  + \mathcal{L}_\texttt{pros.} + \beta \mathcal{L}_\texttt{KL}.
\end{equation}
The first three loss terms in Eq.\eqref{eq:loss} denote the duration loss, VQ classification loss and auxiliary loss in VQTTS~\cite{VQTTS}, and $\mathcal L_{\texttt{KL}}$ refers to the KL-divergence loss brought by the VAE encoder.
Here $\beta$ denotes the KL-divergence annealing weight, which controls the optimization strength over the KL-loss to maintain a tradeoff of avoiding KL vanishing~\cite{KL-An} and overfitting. 
During inference, we scale the predicted durations by a coefficient according to the average value of phone durations in reference speech, to make the speaking rate controlled as well.
Together with the Global Style VAE Encoder and LLM Prompt Selector, the TTS pipeline is empowered with the ability to synthesize high-quality speech with the style either specified by users or inferred from texts.
\vspace{-0.2cm}
\section{EXPERIMENT}
\subsection{Experimental Setup}
The dataset used in this paper consisted of about 60 hours (30k utterances) of recorded Mandarin artistic storytelling shows from a female performer in 16kHz.
The dataset came up with a human-revised transcript after ASR and corresponding aligned durations.
As an art performance speech dataset, the characteristics of high expressiveness are revealed in many aspects, such as great dynamic in pitch and loudness, rich emotional and contextual styles, and detailed artistic techniques including onomatopoeia, cadence, etc.
As mentioned above, our style prompt annotations only cover 200-300 utterances, which is a plug-and-play process and satisfies out-of-domain style prompt requirements via LLM.


In the TTS pipeline, the VQ features were extracted by a publicly available k-means-based vq-wav2vec\footnote{\url{https://github.com/pytorch/fairseq/tree/main/examples/wav2vec}}.
The mel-spectrograms and 3-dimensional auxiliary features were extracted by Kaldi~\cite{Kaldi} in 12.5ms frameshift and 50ms frame length. 
During training, we fixed KL annealing weight $\beta$ to $10^{-6}$ and VAE latent dimensions 8. 
The text2vec and vec2wav were optimized by Adam Optimizer with Noam scheduler on a GTX 2080ti GPU. 

To demonstrate the practicality of our system, the evaluations comprise three parts. The first two parts will separately discuss the audio synthesis pipeline and the LLM-annotation strategy to showcase their capabilities. The final part will assess the overall performance of FS-TTS, encompassing the entire process from user prompts to final audio outputs in style selection and inference. Collectively, these evaluations illustrate a robust, practical, and user-friendly expressive TTS.

\subsection{Style Control with Specified References}
\label{exp1}
In this section, we only evaluate the controllability of the acoustic model with the Global Style VAE Encoder.
We conditioned the VAE Encoder with references that already existed in the annotated set, and measured the extent to which the synthetic speech could convey the specified style.
We conducted subjective listening tests where raters gave mean opinion scores (MOS) based on the naturalness of synthetic utterances, and relevance MOS (RMOS) based on how much the synthetic style resembled the reference.
Both MOS and RMOS were rated in a range of 1-5.



\begin{table}[h]
\vspace{-0.4cm}
\centering
\caption{MOS and RMOS comparison for specified references. Here MOS measures the naturalness of synthetic speech, while RMOS measures the style relevance to the specified reference.}
\begin{tabular}{@{}lccc@{}}
\toprule
\multicolumn{1}{c}{\textbf{Method}} & \textbf{MOS} & \textbf{RMOS} \\ \midrule
GT (voc.) & 4.48$\pm$0.09 & - \\ \midrule
TTS w/o Style & 4.11$\pm$0.08 & 2.94$\pm$0.10 \\
TTS w/ Specified Style 
 & 4.03$\pm$0.11 & \textbf{3.44$\pm$0.10} \\ \bottomrule
\end{tabular}
\label{tab:ExpMOS}
\end{table}

The results are shown in Table \ref{tab:ExpMOS}, where ``GT (voc.)" means the resynthesized ground truth recordings, ``TTS w/o Style" means the baseline acoustic model without Style Encoder, and ``TTS w/ Specified Style" stands for the acoustic model conditioned on specified references.
It is obvious that the conditioned ones outperform the others in style similarity and have similar naturalness with the baseline.
This proves that the acoustic model with Global Style VAE Encoder is effective for style control without significant sacrifice in naturalness, paving the way for freestyle control scenarios.

\vspace{-0.2cm}
\subsection{LLM-Annotation Ablation}
In this part, we focus on the ablation of the number of annotated utterances and evaluate LLM's retrieval performance. Our experiments were devised as follows:
\begin{enumerate}[leftmargin=*,noitemsep, topsep=5pt]
    \item We started with 500 external style descriptions or speech transcripts that were not part of our annotations. Using the Language Model (LLM), we matched these descriptions to the key-value pairs in the dictionary we annotated in \ref{2.2}.
    \item For our questionnaire, we randomly selected 20 items from the matching list for each participant. Participants were asked a simple yes-or-no question: ``Does this annotation correspond to the prompt's description or style?" A ``yes'' indicated a match, while a ``no'' indicated a mismatch.
    \item We calculated the \textbf{hit rate}, which is the percentage of participants who chose ``yes" from among all the questionnaires, for each configuration and the number of annotated utterances.
\end{enumerate}

For LLM Prompt Selector, two configurations were used:
GPT-3.5-turbo-16k and GPT-4~\cite{GPT4}. 
For comparison, we also included a RoBERTa~\cite{RoBERTa} pre-trained in Chinese and fine-tuned by SimCSE~\cite{SimCSE} from HuggingFace\footnote{\url{https://huggingface.co/cyclone/simcse-chinese-roberta-wwm-ext}}. 
For RoBERTa, cosine similarities were computed between prompts and annotations as a selection baseline. 
Both GPT models released by OpenAI were invoked by API requests using the task description we mentioned in \ref{2.2}.

\begin{figure}[htb]
    \centering
    \includegraphics[width=0.75\linewidth]{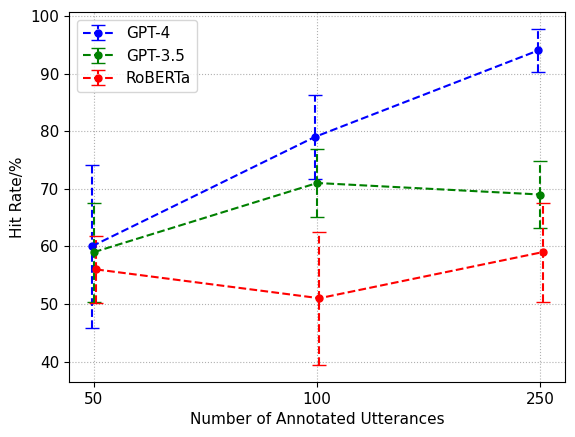}
    \vspace{-0.4cm}
    \caption{The ablation of different LLMs and amounts of annotations. Error bars represent standard deviation.}
    \label{ablation}
    \vspace{-0.2cm}
\end{figure}

In Fig~\ref{ablation}, we can see that GPT-4 performed the best for this task.
As the number of annotations grew to 250, it could match the majority of unseen prompts to suitable annotation labels. 
Thus we can use less than 300 annotations to handle various prompts via LLM. 
In the following freestyle control experiments, we adopted the annotations returned by GPT-4 with 250 annotated utterances.

\vspace{-0.2cm}
\subsection{Free Style Control with LLM}
In this experiment, the TTS pipeline with Global Style VAE Encoder and LLM Prompt Selector worked together to form our FS-TTS system,
and we evaluated the end-to-end performance of it.
For the two prompting scenarios we have mentioned, different designs of experiments were adopted:
  \vspace{-0.1cm}

\subsubsection{Style Selection}
    In this case, the provided prompts were unseen style descriptions, like ``I whispered conspiracy".
    This requires the LLM to compare the similarities between annotated styles and prompts to retrieve the most suitable annotation, and finally guide the TTS pipeline to synthesize speeches corresponding to the descriptions. For evaluation, we conducted an RMOS test similar to Table \ref{tab:ExpMOS}, but this time our reference was only the style description, not reference speech. 
    Participants listened to the test items generated from 60 descriptions and rated them based on the similarity between the style description and the speech. 
    The compared systems included the baseline ``TTS w/o Style", and also a TTS with reference encoder but conditioned by random samples.
    
    The RMOS results are shown in Table \ref{selection}.
    It can be seen that LLM-retrieved style prompts led to an RMOS score that outperformed random ones by a large margin, which demonstrates that FS-TTS can effectively determine reference conditions for synthesis, and produce speech with highly-matched styles.
\begin{table}[h!t]
\vspace{-0.4cm}
   \centering
   \caption{Style similarity RMOS tests in the freestyle selection task. The ``Style Condition" column specifies how the references were chosen for conditioning the reference encoder.}
      \vspace{0.2cm}
   \label{selection}
\begin{tabular}{@{}ccc@{}}
\toprule
\textbf{Method} & \textbf{Style Condition} & \textbf{RMOS}          \\ \midrule
TTS w/o Style & - & 2.70$\pm$0.12 \\ \midrule
\multirow{2}{*}{TTS w/ Style} & Random    & 2.75$\pm$0.12 \\
& LLM-retrieved (FS-TTS) & \textbf{3.33$\pm$0.13} \\ \bottomrule
\end{tabular}
      \vspace{-0.2cm}

\end{table}

\subsubsection{Style Inference}
    In this case, the provided prompts were new speech texts, like ``Hurry up! Somebody call an ambulance!". This requires the LLM to infer the appropriate style from the text. The LLM then matched the inferred style to retrieve the most suitable annotation, and guided the TTS pipeline to synthesize speech that aligned with the annotation and the phoneme sequences of the texts. 
    Since the styles were implicit within the texts, we conducted a pairwise preference test to determine whether the expressiveness of FS-TTS was preferred by humans against a baseline that did not use the reference encoder for style modeling. 
    Listeners were asked to choose the speech sample with the more suitable style according to the speech transcription sampled from 60 speech texts, or "Cannot Decide" when there was no clear preference.

    The results of the preference test can be found in Table \ref{inference}, where FS-TTS with inferred style was preferred in over two-thirds of the cases.
    This verifies that the proposed FS-TTS significantly outperforms the baseline without inferred style, confirming our system's ability to accurately infer speech styles from texts and synthesize speech that aligns well with the style conveyed.
\begin{table}[h!t]
\vspace{-0.4cm}

   \centering
   \caption{ Preference test in freestyle inference task. }
      \vspace{0.2cm}
   \label{inference}
		\begin{tabular}{c  c c  } \toprule
	\textbf{w/ Inferred Style } &    \textbf{No Preference} &\textbf{w/o Inferred Style} \\ \midrule
         66.7\%& 5.8\% &27.5\% \\
              \bottomrule
    \end{tabular}
    \vspace{-0.4cm}
\end{table}
\vspace{-0.2cm}
\section{CONCLUSION}
In this paper, we proposed FreeStyleTTS, an expressive TTS system driven by natural language prompts using few manual annotations. 
FreeStyleTTS utilizes the semantic inference ability of LLMs to transform the natural language style control task in TTS into a style retrieval task.
The easy workload and direct prompting stage make it a step closer to the ideal user-controllable expressive TTS. Our evaluations prove that diverse out-of-domain style descriptions or speech texts can be matched to our annotation via our LLM prompt selector and guide the TTS pipeline to synthesize speeches with correct and suitable styles. 
Future research may focus on refining style control for non-verbal vocalizations like laughter and sobbing, enhancing style prompt matching to multiple references, and optimizing prompt configurations to expand the applicability of our expressive TTS system.


\vfill\pagebreak
\bibliographystyle{IEEEtran}
\bibliography{refs}

\end{document}